\documentstyle[12pt]{article}
\makeatletter
\@addtoreset{equation}{section}

\makeatother
\def\G1{\displaystyle \mathop{G}}
\def\e1{\displaystyle \mathop{e}}
\def\O1{\displaystyle \mathop{O}}
\def\T{\displaystyle \mathop{T}}
\def\R{\displaystyle \mathop{R}}
\def\a{\displaystyle \mathop{a}}
\begin{document}
\begin{center}
\Large {\bf GRAVITATION GAUGE GROUP}\\
\vskip 1truecm
{\normalsize G.T.Ter-Kazarian\footnote
{E-mail address:gago@bao.sci.am}}\\
{\small Byurakan Astrophysical Observatory, Armenia 378433}\\
{\small June 3, 1996}\\
\end{center}
\begin{abstract}
Suggested theory involves a drastic revision of a role of local internal
symmetries in physical concept of curved geometry.
Under the reflection of fields and their dynamics from Minkowski
to Riemannian space a standard gauge principle of local internal symmetries 
is generalized. The gravitation gauge group is proposed, which is generated 
by hidden local internal symmetries.
In all circumstances, it seemed to be of the most importance for understanding 
of physical nature of gravity. The most promising aspect in our approach 
so far is the fact that the energy~-momentum conservation laws of 
gravitational interacting fields are formulated quite naturally 
by exploiting  whole advantage of auxiliary 
{\em shadow fields} on flat {\em shadow space}.
The developed mechanism enables one to infer Einstein~'s 
equation of gravitation, but only with strong difference from
Einstein~'s theory at the vital point of well~-defined energy~-momentum
tensor of gravitational field and conservation laws.
The gravitational interaction as well as general distortion of
manifold $G(2.2.3)$ with  hidden group  $U^{loc}(1)$ was considered.
\end{abstract}
\section {Introduction}
\label {int}
In spite of unrivaled by its simplicity and  beautiful features, Einstein~'s
classical theory of gravitation clashes with some basic principle of field 
theory. This state of affairs has not much changed up to present and proposed 
abundant models of gravitation are not conductive to provide unartificial
and unique recipe for resolving controversial problems of energy-momentum 
conservation laws of gravitational interacting fields, localization of energy 
of gravitation waves and also severe problems involved in quantum gravity. 
It may seem foolhardy to think the rules governing such issues
in scope of the theory accounting for gravitation entirely in terms of
intricated Riemannian geometry. The difficulties associated with this step 
are notorious, however, these difficulties are technical. In the main,
they stem from the fact that Riemannian geometry, in general, has not
admitted a group of isometries. So, the Poincare 
transformations no longer act as isometries. For example, it is
not possible to define energy~-momentum as Noether currents related to exact 
symmetries and so on. On the other hand, the concepts of local internal 
symmetries and gauge fields [1-3] have became a powerful tool and successfully
utilized for the study of electroweak and strong interactions. The intensive 
attempts have been made for constructing a gauge theory of gravitation.
But they are complicated sure enough and not generally acceptable.
Since none of the solutions of the problems of determination of the gauge
group of gravitation and Lagrangian of gravitational field proposed
up to the present seems to be wholly convincing,
one might gain insight into some of unknown features of phenomena of
local internal symmetries and gravity by investigating a new approaches in 
hope of resolving of such issues.
Effecting a reconciliation it is the purpose of present article to explore a 
number of fascinating features of {\em generating the gauge group of gravitation
by hidden local internal symmetry.} This is the guiding principle framing our
discussion. In line with this we feel it of the most importance to 
generalize standard 
gauge principle of local internal symmetries. While, a second trend
emerged as a formalism of reflection of physical fields
and their dynamics from Minkowski space to Riemannian. There is an attempt 
to supply some of the answers in a concise form to the crucial problems of
gravitation and to trace some of the major currents of thoughts under a
novel view-point. Exploiting a whole advantage of the field theory 
in terms of flat space, a particular emphasis will be placed just on 
the formalism of reflection.
This is not a final report on a closed subject, but it is hoped that
suggested theory will serve as useful introduction and that it will 
thereby add the knowledge on the role of local internal symmetries 
in physical concept of curved geometry. Of course, much remains to be done
before one can determine whether this approach can ever contribute to the
larger goal of gaining new insight into the theory of gravity.
\section {Formulation of Principle}
\label{Prin}
In standard picture, suppose a massless gauge field 
${\bf B}_{l}(x_{f})=T^{a}B_{l}^{a}(x_{f})$
with the values in Lie algebra of group G is a local form of expression of 
connection in principle bundle with a structure 
group G. Collection of
matter fields are defined as the sections of vector bundles 
associated with G by reflection $\Phi_{f}:M^{4}\rightarrow E$ that
$p\,\Phi_{f}(x_{f})=x_{f}$, where 
$x_{f} \in M^{4}$ is a space-time point of Minkowski flat space 
specified by index $({}_{f})$. The $\Phi_{f}$ is a column vector denoting 
particular component of field taking values in standard fiber $F_{x_{f}}$
upon $x_{f}$ : $p^{-1}(U^{(f)})=U^{(f)}\times F_{x_{f}}$, where 
$U^{(f)}$ is a region of base of principle bundle 
upon which an expansion into direct product $p^{-1}(U^{(f)})=
U^{(f)}\times G$ is defined. The various suffixes of 
$\Phi_{f}$ are left implicit.  The fiber is Hilbert vector space 
on which a linear representation $U_{f}(x_{f})=
\exp \left( -T^{a}\theta_{f}^{a}(x_{f}) \right )$ of group G with 
structure constants $C^{abc}$ is given. This space regarded as Lie algebra 
of group G upon which Lie algebra acts according to 
law of adjoint representation: ${\bf B}\, \leftrightarrow \, ad \, 
{\bf B}\: \, \Phi_{f} \, \rightarrow [{\bf B}\, ,\Phi_{f}]$.
To facilitate writing,  
we shall consider, in the main, the most important fields of spin 
$0,\displaystyle\frac{1}{2}, 1$. 
But developed method may be readily extended to the field
of arbitrary spin $s$, since latter will be treated as a system of 2$s$
fermions of half~-integral spin.
In order to generalize standard gauge principle below we proceed 
with preliminary discussion.\\
a. As a starting point we shall assume that under gauge field ${\bf B}_{f}$
the basis vectors $e^{l}_{f}$ transformed into four vector
fields 
\begin{equation}
\label {R21}
\hat{e}^{\mu}(\kappa {\bf B}_{f})=
\hat{D}^{\mu}_{l}(\kappa {\bf B}_{f})e^{l}_{f},
\end{equation}
where $\hat{D}^{\mu}_{l}$ are real~-valued matrix~-functions of 
${\bf B}_{f}$, $\kappa$ is universal coupling constant by which 
gravitational constant will be expressed (see eq.(4.7)). 
The double occurrence of dummy indices, as usual, will be taken to 
denote a summation extended over their all values. Explicit form of 
$\hat{D}^{\mu}_{l}$ will not concern us here, which must be defined 
under concrete physical considerations [4-8] (sec. 5,6). However, 
two of the most common restrictions will be placed upon these functions.
As far as all known interactions are studied on the base of commutative
geometry, then, first of all, the functions $\hat{D}^{\mu}_{l}$ will be 
diagonalized at given
$(\mu, l)$. As a real~-valued function of Hermitian matrix ${\bf B}_{f}$,
each of them is Hermitian too and may be diagonalized by proper unitary 
matrix. Hence
$D^{\mu}_{l}(\kappa {\bf B}_{f})=diag(\lambda^{\mu}_{l1},
\lambda^{\mu}_{l2},\lambda^{\mu}_{l3},\lambda^{\mu}_{l4}),$
provided with eigen~-values $\lambda^{\mu}_{li}$ as the roots of polynomial
characteristic equation
\begin{equation}
\label {R22}
C(\lambda^{\mu}_{l})=det(\lambda^{\mu}_{l}I-\hat{D}^{\mu}_{l})=0.
\end{equation}
Thus
\begin{equation}
\label {R23}
e^{\mu}(\kappa {\bf B}_{f})=
D^{\mu}_{l}(\kappa {\bf B}_{f})e^{l}_{f}
\end{equation}
lead to commutative geometry. Consequently, if $e_{\nu}(\kappa {\bf B}_{f})$
denotes the inverse matrix~-vector $\|D\| \neq 0$, then
$<e^{\mu},e_{\nu}>=D^{\mu}_{l}D_{\nu}^{l}=\delta_{\nu}^{\mu}, \quad
D^{\mu}_{l}D_{\mu}^{k}=\delta_{l}^{k}.$
A bilinear form on vector fields of sections $\tau$ of tangent bundle of
$R^{4}: \quad 
\hat{g}:\tau \otimes \tau \rightarrow C^{\infty} (R^{4}),$
namely the metric $(g_{\mu \nu})$, is a section of conjugate vector bundle 
$S^{2}\tau^{*}$ 
(symmetric part of tensor degree) with corresponding components in basis
$e_{\mu}(\kappa {\bf B}_{f})$
\begin{equation}
\label {R25}
g_{\mu \nu}=<e_{\mu},e_{\nu}>=<e_{\nu},e_{\mu}>=g_{\nu \mu},\quad
e_{\mu}=g_{\mu \nu}e^{\nu}.
\end{equation}
In holonomic basis
$\hat{g}=g_{\mu \nu}dx^{\mu}\otimes dx^{\nu}.$
One now has to impose a second restriction on the functions $D^{\mu}_{l}$ by
placing stringent condition upon the tensor $\omega^{m}_{l}(x_{f})$ implying
\begin{equation}
\label {R26}
\partial^{f}_{l}D^{\mu}_{k}(\kappa {\bf B}_{f})=
\partial^{f}_{k}\left( \omega^{m}_{l}(x_{f})
D^{\mu}_{m}(\kappa {\bf B}_{f})\right),
\end{equation}
where $\partial^{f}_{l}={\displaystyle \frac{\partial}{\partial x_{f}^{l}}}$,
that the $\omega=\omega^{l}_{l}$ should be Lorentz scalar 
function of the trace of curvature form $\Omega$ of connection 
${\bf B}_{l}$ with the values in Lie algebra of group G. That is
\begin{equation}
\label {R27}
\omega = \omega (tr\, \Omega) = \omega (d\,tr \,{\bf B}), 
\end{equation}
provided
\begin{equation}
\label {R28}
\begin{array}{l}
\Omega = \displaystyle{\sum_{l < k}} {\bf F}_{lk}^{f}\,dx^{l}_{f}
\wedge dx^{k}_{f}, \quad {\bf B}={\bf B}_{l}^{f}dx^{l}_{f},\\
{\bf F}_{lk}^{f}=\partial_{l}^{f}{\bf B}_{k}-\partial_{k}^{f}{\bf B}_{l}
-ig[{\bf B}_{l},{\bf B}_{k}].
\end{array}
\end{equation}
There up on a functional $\omega$ has a null variational derivative
$\displaystyle \frac{\delta \omega (tr \,\Omega )}{\delta {\bf B}_{l}} = 0 $
at local variations of connection ${\bf B}_{l} \rightarrow {\bf B}_{l}+
\delta {\bf B}_{l}$, namely, $\omega$ is invariant under Lorentz($\Lambda$) and
G-gauge  groups
$\omega = inv(\Lambda, G).$\\
b. Constructing a diffeomorphism
$x^{\mu}(x^{l}_{f}):M^{4}\rightarrow R^{4},$
the holonomic functions $x^{\mu}(x^{l}_{f})$ satisfy defining
relation
\begin{equation}
\label {R29}
e_{\mu}\psi^{\mu}_{l}=e^{f}_{l} + \chi^{f}_{l}({\bf B}_{f}),
\end{equation}
where
\begin{equation}
\label {R210}
\chi^{f}_{l}({\bf B}_{f})=e_{\mu}\chi^{\mu}_{l}=
-\frac{1}{2}e_{\mu}\int_{0}^{x_{f}}
(\partial^{f}_{k}D^{\mu}_{l}-\partial^{f}_{l}D^{\mu}_{k})dx^{k}_{f}.
\end{equation}
The following 
notational conventions will be used throughout: $\psi^{\mu}_{l}=
\partial^{f}_{l}x^{\mu}$, where the indices $\mu,\nu,\lambda,\tau,\sigma,
\kappa$ stand for variables in $R^{4}$, when $l,k,m,n,i,j$ refer to 
$M^{4}$. A closer examination of eq.(2.9) shows that 
covector $\chi^{f}_{l}({\bf B}_{f})$   realizes coordinates $x^{\mu}$ 
by providing a criteria of integration
\begin{equation}
\label {R211}
\partial^{f}_{k}\psi^{\mu}_{l}=\partial^{f}_{l}\psi^{\mu}_{k}
\end{equation}
and undegeneration $\| \psi \| \neq 0$ [9,10]. Due to eq.(2.6) and 
eq.(2.9), one has Lorentz scalar gauge
invariant functions
\begin{equation}
\label {R212}
\chi=<e^{l}_{f},\chi^{f}_{l}>=\displaystyle {\frac{\omega}{2}}-2,\quad
S=\displaystyle {\frac{1}{4}}\psi_{\mu}^{l}D^{\mu}_{l}=
\displaystyle {\frac{1}{2}(1 +
\displaystyle {\frac{\omega}{4}}}).
\end{equation}
So, out of a set of arbitrary curvilinear coordinates in $R^{4}$ the
{\em real~-curvilinear} coordinates may be distinguished, which satisfy
eq.(2.8) under all possible Lorentz and gauge transformations. There is
a single~-valued conformity between corresponding tensors with various
suffixes on $R^{4}$ and $M^{4}$. While, each co- or contra~-variant index
transformed incorporating respectively with functions $\psi^{\mu}_{l}$
or $\psi_{\mu}^{l}$. Each transformation of real~-curvilinear coordinates
$x'^{\mu'} \rightarrow x^{\mu}$ was generated by some Lorentz and gauge
transformations
\begin{equation}
\label {R212}
\frac{\partial x'^{\mu'}}{\partial x^{\mu}}=\psi^{\mu'}_{l'}(B'_{f})
\psi_{\mu}^{l}(B_{f})\Lambda^{l'}_{l}.
\end{equation}
There would then exist preferred systems and group of transformations of
real~-curvilinear coordinates in $R^{4}$. The wider group of transformations
of arbitrary curvilinear coordinates in $R^{4}$ would then be of no importance 
for the field dynamics.
If an inverse function $\psi_{\mu}^{l}$ meets condition 
\begin{equation}
\label {R213}
\frac{\partial \psi_{\mu}^{l}}{dx^{\nu}}\neq\Gamma^{\lambda}_{\mu\nu}
\psi_{\lambda}^{l},
\end{equation}
where $\Gamma^{\lambda}_{\mu\nu}$ is the usual Christoffel symbol 
agreed with a 
metric $g_{\mu\nu}$, then a curvature of $R^{4}$ is
not vanished [11].\\
In pursuing the original problem further we are led to the principle point
of drastic change of standard gauge scheme
to assume hereafter that {\em single~-valued, smooth, double-sided 
reflection of fields 
$\Phi_{f}\rightarrow \Phi(\Phi_{f}): F_{x_{f}}\rightarrow F_{x}$, 
namely $\Phi_{f}^{l\cdots m}(x_{f})$, takes place under local 
group G}, where $\Phi_{f}\subset F_{x_{f}}$, $\Phi\subset F_{x}$, $F_{x}$
is the fiber upon $x:p^{-1}(U)=U\times F_{x}$, $U$ is the region of base 
$R^{4}$.
The tensor suffixes were only put forth in illustration of
a point at issue. Then
\begin{equation}
\label {R214}
\Phi^{\mu\cdots \delta}(x)=\psi^{\mu}_{l}\cdots 
\psi^{\delta}_{m}R({\bf B}_{f})\Phi_{f}^{l\cdots m}(x_{f})\equiv
{(R_{\psi})}^{\mu \cdots \delta}_{l\cdots m}\Phi_{f}^{l\cdots m}(x_{f}),
\end{equation}
where $R({\bf B}_{f})$ is a reflection matrix.
The idea of {\em general gauge principle} may be framed into requirement
of {\em invariance of physical system of fields $\Phi(x)$ under the finite 
local gauge transformations $U_{R}=R'_{\psi}U_{f}R^{+}_{\psi}$ of the 
Lie group of gravitation $G_{R}$ generated by G, where $R'_{\psi}=R_{\psi}
({\bf B}'_{f})$ if gauge field ${\bf B}_{f}(x_{f})$ was transformed under 
G in standard form}. While the corresponding 
transformations of fields $\Phi(x)$ and their covariant derivatives 
are written 
\begin{equation}
\label {R216}
\begin{array}{l}
\Phi'(x)=U_{R}(x)\Phi(x),
\\
\left( g^{\mu}(x)\nabla_{\mu}\Phi(x)\right)' =U_{R}(x) \left( 
g^{\mu}(x)\nabla_{\mu}\Phi(x)\right).
\end{array}
\end{equation}
The solution of eq.(2.15) may be readily obtained as the 
reflection of covariant derivatives 
\begin{equation}
\label {R217}
g^{\nu}(x)\nabla_{\nu}\Phi^{\mu\cdots \delta}(x)=S(B_{f})
\psi^{\mu}_{l}\cdots \psi^{\delta}_{m}R({\bf B}_{f})\gamma^{k}D_{k}
\Phi_{f}^{l\cdots m}(x_{f}),
\end{equation}
where $D_{l}=\partial_{l}^{f}- ig {\bf B}_{l}(x_{f})$ , 
$S(B_{f})$ is gauge invariant Lorentz scalar, $\nabla_{\mu}$ is 
covariant derivative in $R^{4}$ :
$\nabla_{\mu}=\partial_{\mu}+\Gamma_{\mu}$, provided with connection [11]
$\Gamma_{\mu}(x)=\displaystyle\frac{1}{2}
\Sigma^{\alpha\beta}V^{\nu}_{\alpha}(x)
\partial_{\mu}V_{\beta\nu}(x),$
the $\Sigma^{\alpha\beta}$ are generators of Lorentz group, 
$V^{\mu}_{\alpha}(x)$ are the components of affine tetrad vectors 
$e^{\alpha}$ in used coordinate net $x^{\mu}$: 
$V^{\mu}_{\alpha}(x)=<e^{\mu},e_{\alpha}>$; one has $g^{\mu}(x)\Rightarrow 
e^{\mu}(x)$ and $\gamma^{l}\Rightarrow e^{l}_{f}$ 
for fields $(s=0,1)$; but $g^{\mu}(x)=
V^{\mu}_{\alpha}(x)\gamma^{\alpha}$ for spinor field $(s=\displaystyle
\frac{1}{2})$, where $\gamma^{\alpha}$ are Dirac's matrices.
Since the fields
$\Phi_{f}(x_{f})$ no longer hold, the reflected ones $\Phi(x)$ will be
regarded as the real physical fields. But a conformity eq.(2.14) and eq(2.16)
enables $\Phi_{f}(x_{f})$ to serve as an auxiliary 
{\em shadow fields} on {\em shadow flat space $M^{4}$}. These notions
arise basically  from the most important fact that a 
Lagrangian $L(x)$ of fields $\Phi(x)$
may be obtained under the reflection from a Lagrangian $L_{f}(x_{f})$ of 
corresponding shadow fields and vice versa. Certainly, the $L(x)$ is
also an invariant under the wider group of arbitrary curvilinear 
transformations $x \rightarrow x'$ in $R^{4}$
\begin{equation}
\label {R217}
\left. J_{\psi} L(x)\right|_{inv(G_{R}; \, x\rightarrow x')}=
\left. L_{f}(x_{f})\right|_{inv(\Lambda; \, G)},
\end{equation}
where $J_{\psi}= \| \psi\|\,\surd =\left(1+2\|<e^{f}_{l},\chi^{f}_{k}>\|
+ \| <\chi^{f}_{l},\chi^{f}_{k}>\|\right)^{1/2}$ (see eq.(4.1)).
While, {\em the internal gauge symmetry $G$ remained hidden symmetry,
since it screened by gauge group of gravitation $G_{R}$}.
We note that the tetrad $e^{\alpha}$ and basis $e^{l}_{f}$
vectors meet a condition
\begin{equation}
\label {R219}
\begin{array}{l}
\rho^{\alpha}_{l}(x,x_{f})=<e^{\alpha},e_{l}^{f}>=V^{\alpha}_{\mu}(x)
D^{\mu}_{l}(\kappa {\bf B}_{f}), \\
<e^{\alpha}_{f},e^{f}_{\beta}>=<e^{\alpha},e_{\beta}>=
\delta^{\alpha}_{\beta}.
\end{array}
\end{equation}
So, Minkowskian metric is written
$\eta^{\alpha\beta}=diag(1,-1,-1,-1)= 
<e^{\alpha}_{f},e_{f}^{\beta}>=<e^{\alpha},e^{\beta}>$.
\section {Reflection Matrix}
\label {ref}
One interesting offshoot of general gauge principle is a formalism of
reflection.
A straightforward calculation for fields $s=0,1$ 
gives the explicit unitary reflection matrix
\begin{equation}
\label {R31}
R(x,x_{f})=R_{f}(x_{f})R_{g}(x)=\exp\left( 
-i\Theta_{f}(x_{f})-
\Theta_{g}(x)\right),
\end{equation}
provided
\begin{equation}
\label {R32}
\Theta_{f}(x_{f})=
g\int_{0}^{x_{f}}{\bf B}_{l}(x_{f})dx_{f}^{l},
\quad \Theta_{g}(x)=
\int_{0}^{x}\left[ R^{+}_{f} \Gamma_{\mu} R_{f}+
\psi^{-1}\partial_{\mu} \psi\right]dx^{\mu},
\quad \psi \equiv \left(\psi^{\mu}_{l}\right),
\end{equation}
where $\Theta_{g}=0$ for scalar field and $\Theta_{g}+\Theta_{g}^{+}=0$
for vector field, because of $\Gamma_{\mu}+\Gamma_{\mu}^{+}=0$ and 
$\partial_{\mu} (\psi^{-1}\psi)=0$.
The function $S(B_{f})$ has a form eq.(2.11), since
$S(B_{f})=\displaystyle \frac{1}{4}R^{+}(\psi^{l}_{\mu}D_{l}^{\mu})R
=\displaystyle \frac{1}{4}\psi^{l}_{\mu}D_{l}^{\mu}.$
The infinitesimal gauge transformation 
$U_{f}\approx 1-iT^{a}\theta^{a}_{f}(x_{f})$ yields
\begin{equation}
\label {R33}
U_{R}=\exp \left( igC^{abc}T^{a}\int_{0}^{x_{f}}\theta^{b}_{f}B^{c}_{l}
dx^{l}_{f} + \Theta'_{g}-\Theta_{g}\right),
\end{equation}
where the infinitesimal transformation ${B'}^{a}_{l}=B^{a}_{l} + 
C^{abc}\theta^{b}_{f}B^{c}_{l} - \displaystyle\frac{1}{g}\partial^{f}_{l}
\theta^{a}_{f}$ was used.
For example, a Lagrangian of {\em isospinor-scalar} shadow field 
$\varphi_{f}$ is in the form
\begin{equation}
\label {R34}
\begin{array}{l}
L_{f}(x_{f})=J_{\psi}L(x)=J_{\psi}\left[ (e^{\mu}\nabla_{\mu}\varphi)^{+}
(e^{\mu}\nabla_{\mu}\varphi)- m^{2}\varphi^{+}\varphi \right]=\\
=J_{\psi}\left[ S(B_{f})^{2}(D_{l}\varphi_{f})^{+}(D_{l}\varphi_{f})-
m^{2}\varphi_{f}^{+}\varphi_{f}\right].
\end{array}
\end{equation}
A Lagrangian of {\em isospinor-vector} Maxwell~'s shadow field arises in a 
straightforward manner
\begin{equation}
\label {R35}
\begin{array}{l}
L_{f}(x_{f})=J_{\psi}L(x)=\\
=-\displaystyle\frac{1}{4}J_{\psi}F_{\mu\nu}^{+}F^{\mu\nu}=-
\displaystyle\frac{1}{4}J_{\psi}S(B_{f})^{2}
(F^{(f)}_{\mu\nu})^{+}F^{\mu\nu}_{(f)},
\end{array}
\end{equation}
provided
\begin{equation}
\label {R36}
F_{\mu\nu}=\nabla_{\mu}A_{\nu}-\nabla_{\nu}A_{\mu},\quad
F^{(f)}_{\mu\nu}=(\psi^{k}_{\nu}D^{f}_{\mu}-\psi^{k}_{\mu}D^{f}_{\nu})
A^{f}_{k}, \quad D^{f}_{\mu}=D^{l}_{\mu}D_{l},
\end{equation}
with an additional gauge violating term
\begin{equation}
\label {R37}
\begin{array}{l}
L^{f}_{G}(x_{f})=J_{\psi}L_{G}(x)=-\displaystyle \frac{1}{2}\zeta^{-1}_{0}
J_{\psi}(\nabla_{\mu}A^{\mu})^{+}
\nabla_{\nu}A^{\nu}=\\
=-\displaystyle \frac{1}{2}\zeta^{-1}_{0}J_{\psi}S(B_{f})^{2}
(\lambda^{l}_{m}D_{l}A_{f}^{m})^{+}\lambda^{k}_{n}D_{k}A_{f}^{n},
\quad \lambda^{l}_{m}=\displaystyle \frac{1}{2}(\delta^{l}_{m} +
\omega^{l}_{m}),
\end{array}
\end{equation}
where $\zeta^{-1}_{0}$ is gauge fixation parameter. Finally, a Lagrangian
of {\em isospinor-ghost} fields is written
\begin{equation}
\label {R38}
L^{f}_{gh}(x_{f})=J_{\psi}L_{gh}(x)=J_{\psi}
<(e^{\mu}\partial_{\mu}C)^{+},e^{\nu}\partial_{\nu}C>=
J_{\psi}S(B_{f})^{2}(D_{l}C_{f})^{+}D_{l}C_{f}.
\end{equation}
Continuing along this line we come to a discussion of the reflection
of spinor field $\Psi_{f}(x_{f}) (s=\displaystyle\frac{1}{2})$ 
as a solution of eq.(2.15)
\begin{equation}
\label {R39}
\begin{array}{l}
\Psi(x)=R({\bf B}_{f})\Psi_{f}(x_{f}), \quad
\bar{\Psi}(x)=\bar{\Psi}_{f}(x_{f})\widetilde{R}^{+}({\bf B}_{f}),\\
g^{\mu}(x)\nabla_{\mu}\Psi(x)=S(B_{f})R({\bf B}_{f})
\gamma^{l}D_{l}\Psi_{f}(x_{f}),\\
\left( \nabla_{\mu}\bar{\Psi}(x)\right)g^{\mu}(x)=S(B_{f})
\left( D_{l}\bar{\Psi}_{f}(x_{f})\right)\gamma^{l}
\widetilde{R}^{+}({\bf B}_{f}),
\end{array}
\end{equation}
where $\widetilde{R}=\gamma^{0}R\gamma^{0},\quad \Sigma^{\alpha\beta}=
\displaystyle\frac{1}{4}[\gamma^{\alpha},\gamma^{\beta}], \quad 
\Gamma_{\mu}(x)=\displaystyle\frac{1}{4}
\Delta_{\mu,\alpha\beta}(x)\gamma^{\alpha}\gamma^{\beta}$, 
$\Delta_{\mu,\alpha\beta}(x)$ are Rici's rotation coefficients. 
The reflection matrix $R$ is in 
the form eq.(3.1), provided we make change
\begin{equation}
\label {R310}
\Theta_{g}(x)=\frac{1}{2}\int_{0}^{x} R^{+}_{f}\left\{ 
g^{\mu}\Gamma_{\mu}R_{f}, g_{\nu}dx^{\nu}\right\},
\end{equation}
the $\{,\}$ is an anticommutator. A calculation now gives
\begin{equation}
\label {R311}
S(B_{f})=\frac{1}{8K}\psi^{l}_{\mu}\left\{
\widetilde{R}^{+}g^{\mu}R,
\gamma_{l}\right\}=inv,
\end{equation}
where
\begin{equation}
\label {R312}
\begin{array}{l}
K =\widetilde{R}^{+}R=\widetilde{R}^{+}_{g}R_{g}=\\
=\exp\left( -\displaystyle \frac{1}{2}\int_{0}^{x}\left( \left\{ R^{+}_{f}
\widetilde{\Gamma}^{+}_{\mu}g^{\mu},g_{\nu}dx^{\nu}\right\}R_{f}+
R^{+}_{f}\left\{ g^{\mu}\Gamma_{\mu}R_{f},
g_{\nu}dx^{\nu}\right\}\right)\right), 
\quad \widetilde{\Gamma}^{+}_{\mu}=
\gamma^{0}\Gamma^{+}_{\mu}\gamma^{0}.
\end{array}
\end{equation}
Taking into account that $[R_{f}, g_{\nu}]=0$ and substituting [12]
\begin{equation}
\label {R313}
\widetilde{\Gamma}^{+}_{\mu}g^{\nu} + g^{\nu}\Gamma_{\mu}=
-\nabla_{\mu}g^{\nu}=0
\end{equation}
into eq.(3.12), we get
\begin{equation}
\label {R314}
K = 1.
\end{equation}
Since 
\begin{equation}
\label {R315}
\widetilde{U}_{R}^{+}U_{R}=\widetilde{R}U^{+}_{f}
\widetilde{R'}^{+}R'U_{f}R^{+}=
\widetilde{R}U^{+}_{f}
\widetilde{R}^{+}RU_{f}R^{+},
\end{equation}
where $\widetilde{R'}^{+}R'=\widetilde{R}^{+}R=1$, then
\begin{equation}
\label {R316}
\widetilde{U}_{R}^{+}U_{R}=\gamma^{0}U_{R}^{+}\gamma^{0}U_{R}=1.
\end{equation}
A Lagrangian of {\em isospinor-spinor} shadow field may be written
\begin{equation}
\label {R317}
\begin{array}{l}
L_{f}(x_{f})=J_{\psi} L(x)=\\
=J_{\psi}\left\{ 
\displaystyle {\frac{i}{2}}\left[ 
\bar{\Psi}(x)g^{\mu}(x)\nabla_{\mu}\Psi(x)-
(\nabla_{\mu}\bar{\Psi}(x))g^{\mu}(x)\Psi(x)\right]-
m\bar{\Psi}(x)\Psi(x)\right\}=\\
=J_{\psi}\left\{S(B_{f})
\displaystyle {\frac{i}{2}}\left[ 
\bar{\Psi}_{f}\gamma^{l}D_{l}\Psi_{f}-
(D_{l}\bar{\Psi}_{f})\gamma^{l}\Psi_{f}\right]-
m\bar{\Psi}_{f}\Psi_{f}\right\}.
\end{array}
\end{equation}
In special case if curvature tensor 
$R^{\lambda}_{\mu\nu\tau}=0$, the eq.(2.8) may be satisfied 
globally in $M^{4}$ by putting
\begin{equation}
\label {R318}
\psi^{\mu}_{l}=D^{\mu}_{l}=V^{\mu}_{l}=\frac{\partial x^{\mu}}
{\partial \xi^{l}}, \quad \|D\|\neq 0,\quad \chi^{f}_{l}=0,
\end{equation}
where $\xi^{l}$ are inertial coordinates.
So, $S = J_{\psi}=1$, which means that one simply has constructed 
local G-gauge theory in $M^{4}$ both in curvilinear as well as inertial
coordinates.
\section{Action Principle}
\label{Act}
Field equations may be inferred from an invariant action
\begin{equation}
\label {R41}
S=S_{B_{f}} + S_{\Phi}=\int L_{B_{f}}(x_{f})d^{4}x_{f} +
\int \surd L_{\Phi}(x) d^{4}x.
\end{equation}
A Lagrangian $L_{B_{f}}(x_{f})$ of gauge field ${\bf B}_{l}(x_{f})$
defined on $M^{4}$ is invariant under Lorentz as well as G~-gauge groups. 
But a Lagrangian of the rest of fields $\Phi(x)$ defined on $R^{4}$
is invariant under the gauge group of gravitation $G_{R}$. Consequently, 
the whole action eq.(4.1) is G~-gauge invariant, since $G_{R}$ generated by 
G. Field equations followed at once in terms of Euler-Lagrang variations 
respectively in $M^{4}$ and $R^{4}$  
\begin{equation}
\label {R42}
\begin{array}{l}
\displaystyle \frac{\delta^{f} L_{B_{f}}}{\delta^{f}B^{a}_{l}}=
J_{a}^{l}=-\displaystyle \frac{\delta^{f} L_{\Phi_{f}}}
{\delta^{f} B^{a}_{l}}=-
\displaystyle \frac{\partial g^{\mu\nu}}{\partial B^{a}_{l}}
\displaystyle \frac{\delta (\surd L_{\Phi})}{\delta g^{\mu\nu}}=
\\
=-\displaystyle \frac{1}{2}
\displaystyle \frac{\partial g^{\mu\nu}}{\partial B^{a}_{l}}
\displaystyle \frac{\surd}{\|D\|}
\displaystyle \frac{D_{k\mu}\delta (\surd L_{\Phi})}
{\delta D^{\nu}_{k}}=-\displaystyle \frac{\surd}{2}
\displaystyle \frac{\partial g^{\mu\nu}}{\partial B^{a}_{l}}T_{\mu\nu},\\
\\
\displaystyle \frac{\delta L_{\Phi}}{\delta\Phi}=0,\quad
\displaystyle \frac{\delta L_{\Phi}}{\delta\bar\Phi}=0,
\end{array}
\end{equation}
where $T_{\mu\nu}$ is the energy~-momentum tensor of fields $\Phi (x)$.
Making use of Lagrangian of corresponding shadow fields $\Phi_{f}(x_{f})$,
in generalized sense, one may readily define the energy~-momentum
conservation laws and also exploit whole advantage of field theory in terms
of flat space in order to settle or mitigate the difficulties whenever they
arise including the quantization of gravitation, which will not concern us 
here. Meanwhile, of course, one is free to 
carry out an inverse reflection to $R^{4}$ whenever it will be needed.
To render our discussion here more transparent, below we clarify a relation
between gravitational and coupling constants. So, we consider the theory at 
the limit of Newton~'s non~-relativistic law of gravitation in the case
of static weak gravitational field given by Poisson equation [11]
\begin{equation}
\label {R43}
\nabla^{2}g_{00}=8\pi G\, T_{00}.
\end{equation}
In linear approximation
\begin{equation}
\label {R44}
g_{00}(\kappa {\bf B}_{f})\approx \left(1 + 2\kappa B^{a}_{k}(x_{f})
\theta_{a}^{k}\right),
\quad \theta_{a}^{k}=\left( \frac{\partial D^{0}_{0} (\kappa {\bf B}_{f})}
{\partial \kappa B^{a}_{k}}\right)_{0} = \mbox{const},
\end{equation}
then
\begin{equation}
\label {R45}
\theta_{a}^{k}\nabla^{2}\kappa B^{a}_{k}=4\pi G\,T_{00}.
\end{equation}
Since, the eq.(4.2) must match onto eq.(4.5) at considered limit, then
the right~-hand sides of both equations should be in the same form
\begin{equation}
\label {R46}
\kappa \theta_{a}^{k}
\displaystyle \frac{\delta^{f} L_{B_{f}}}{\delta^{f}B_{a}^{k}}=
\kappa \theta_{a}^{k}
J^{a}_{k}\approx -\displaystyle \frac{\kappa^{2}}{2}
\displaystyle \frac{\theta_{a}^{k}\partial g^{00}}
{\partial (\kappa B_{a}^{k})}T_{00}=
\kappa^{2}(\theta_{a}^{k}\theta^{a}_{k})T_{00}.
\end{equation}
With this final detail carried for one gets
\begin{equation}
\label {R47}
G=\frac{\kappa^{2}}{4\pi \theta_{c}^{2}}, \quad
\theta_{c}^{2}\equiv (\theta_{a}^{k}\theta^{a}_{k}).
\end{equation}
At this point we emphasize that Weinberg~'s argumentation [13],
namely, a prediction of attraction between particle and antiparticle, 
and repulsion between the same kind of particles, which is valid for 
pure vector theory, no longer holds in suggested theory of gravitation.
Although the ${\bf B}_{l}(x_{f})$ is vector gauge field,
the eq.(4.7) just furnished the prove that only gravitational attraction
is existed.
One final observation is worth recording. A fascinating opportunity
has turned out in the case if one utilizes reflected Lagrangian of 
Einstein~'s gravitational field
\begin{equation}
\label {R48}
L_{B_{f}}(x_{f})=J_{\psi}L_{E}(x)=\frac{J_{\psi}\,R}{16\pi G},
\end{equation}
where $R$ is a scalar curvature. Hence, one readily gets the field equation
\begin{equation}
\label {R49}
\left( R^{\mu}_{\lambda} - \frac{1}{2}\delta ^{\mu}_{\lambda} R\right)
D_{\mu}^{l}\frac{\partial D_{l}^{\lambda}}{\partial B^{a}_{k}}=
-8\pi G U_{a}^{k}(\kappa {\bf B}_{f})=-8\pi G T_{\lambda\nu}
D^{\nu l}\frac{\partial D_{l}^{\lambda}}{\partial B^{a}_{k}},
\end{equation}
which obviously leads to Einsten~'s equation.
Of course in these circumstances it is straightforward
to choose $g^{\mu\nu}$ as the characteristic of gravitational field
without referring to gauge field $B^{a}_{l}(x_{f})$.
However, in this case the energy~-momentum tensor of gravitational 
field is well~-defined. At this vital point suggested theory strongly 
differs from Einstein's classical theory. 
In line with this, one should use the real~-curvilinear coordinates,
in which the gravitational field was well~-defined in the sense that
it cannot be destroyed globally by coordinate transformations.
While, taking into account general rules eq.(2.12), the energy~-momentum 
tensor of gravitational field may be readily obtained 
by expressing the energy~-momentum tensor of vector gauge field 
$B^{a}_{l}(x_{f})$ in terms of metric tensor and its derivatives
\begin{equation}
\label {R410}
\begin{array}{l}
T_{\mu\nu}=g_{\mu\lambda}T^{\lambda}_{\nu}=g_{\mu\lambda}\psi^{\lambda}_{k}
\psi^{i}_{\nu}T^{\,k}_{(f)i}=\\
\\
=g_{\mu\lambda}\psi^{\lambda}_{k}\psi^{i}_{\nu}
\left( {\displaystyle \frac{\partial B^{a}_{l}}{\partial g^{\mu'\nu'}}
\psi^{\sigma}_{i}\partial _{\sigma}g^{\mu'\nu'}
\displaystyle \frac{\partial \left(\partial_{\tau}
g^{\lambda'\tau'}\right)}{\partial\left(\partial^{f}_{k}\,B^{a}_{l}
\right)}
\displaystyle \frac{\partial (J_{\psi} L_{E})}{\partial\left(  
\partial_{\tau} g^{\lambda'\tau'}\right)}
-\delta^{k}_{i}J_{\psi} L_{E} } \right) .
\end{array}
\end{equation}
At last, we should note that eq.(4.8) is not the simplest one among gauge
invariant Lagrangians. Moreover, it must be the same in all cases including
eq.(3.18) too. But in last case it contravenes the standard
gauge theory. There is no need to contemplate such a drastic revision of 
physics. So, for our part we prefer G~-gauge invariant Lagrangian 
in terms of $M^{4}$
\begin{equation}
\label {R411}
L_{B_{f}}(x_{f})=-\frac{1}{4}<{\bf F}^{f}_{lk}(x_{f}),
{\bf F}^{lk}_{f}(x_{f})>_{K}, 
\end{equation}
where ${\bf F}_{lk}^{f}(x_{f})$ is in the form eq.(2.7), 
$<,>_{K}$ is the Killing undegenerate form on the Lie algebra of 
group G for adjoint representation. Certainly, an explicit form of functions
$D^{\mu}_{l}(\kappa {\bf a}_{f})$ should be defined.
\section{Gravitation at $G=U^{loc}(1)$}
The gravitational interaction with hidden Abelian local group
$G=U^{loc}(1)=SO^{loc}(2)$ and one-dimensional trivial algebra
$\hat{g}=R^{1}$ was considered in [4-8], wherein the explicit form of 
transformation function $D^{\mu}_{l}$ is defined by making use of principle
bundle $p:E\rightarrow G(2.3)$. It is worthwhile to consider the major points 
of it anew in concise form and make it complete by calculations,
which will be adjusted to fit the outlined here theory. We start with
a very brief recapitulation of structure of flat manifold
$G(2.3)={}^{*}R^{2}\otimes R^{3}=R^{3}_{+}\oplus R^{3}_{-}$ provided with
the basis vectors $e^{0}_{(\lambda \alpha)}=O_{\lambda}\otimes 
\sigma_{\alpha}$, where $<O_{\lambda},O_{\tau}>={}^{*}\delta_{\lambda\tau}=
1-\delta_{\lambda\tau}$, $<\sigma_{\alpha},\sigma_{\beta}>=\delta_{\alpha
\beta}$ $(\lambda,\tau=\pm : \alpha, \beta=1,2,3)$, $\delta$ is Kronecker 
symbol. A bilinear form on vector fields of sections $\tau$ of tangent bundle 
of $G(2.3)$:  $\hat{g}^{0}:\tau\otimes\tau \rightarrow C^{\infty}(G(2.3))$,
namely, the metric $(g^{0}_{(\lambda\alpha)(\tau\beta)})$ is a section of 
conjugate vector bundle with components $\hat{g^{0}}
(e^{0}_{(\lambda \alpha)}, e^{0}_{(\tau\beta)})$ in basis 
$(e^{0}_{(\lambda \alpha)})$. The $G(2.3)$ decomposes into three-dimensional
ordinary $(R_{f}^{3})$ and time $(T_{f}^{3})$ flat spaces $G(2.3)=
R_{f}^{3}\oplus T_{f}^{3}$ with signatures $sgn(R_{f}^{3})=(+++)$ and
$sgn(T_{f}^{3})=(---)$. Since all directions in $T_{f}^{3}$ are
equivalent, then by notion {\em time} one implies the projection of
time-coordinate on fixed arbitrary universal direction in $T_{f}^{3}$.
By this reduction $T_{f}^{3}\rightarrow T_{f}^{1}$ the transition
$G(2.3)\rightarrow M^{4}=R_{f}^{3}\oplus T_{f}^{1}$ may be performed
whenever it will be needed. 
Under massless gauge field $a_{(\lambda\alpha)}(\eta_{f})$ associating with
$U^{loc}(1)$ the basis $e^{0}_{(\lambda \alpha)}$ transformed at a point
$\eta_{f}\in G(2.3)$ according to eq.(2.3)
\begin{equation}
\label{R51}
e_{(\lambda \alpha)}=D^{(\tau\beta)}_{(\lambda \alpha)}e^{0}_{(\tau\beta)}.
\end{equation}
While, the matrix $D$ is in the form $D=C\otimes R$, where the distortion
transformations $O_{(\lambda\alpha)}=C^{\tau}_{(\lambda\alpha)}
O_{\tau}$ and
$\sigma_{(\lambda\alpha)}=R^{\beta}_{(\lambda\alpha)}
O_{\beta}$ are defined. Thereby $C^{\tau}_{(\lambda\alpha)}=
\delta^{\tau}_{\lambda} + \kappa a_{(\lambda\alpha)}{}^{*}
\delta^{\tau}_{\lambda}$,
but $R$ is a matrix of the group $SO(3)$ of all ordinary rotations of 
the planes, each of which involves two arbitrary basis vectors 
of $R^{3}_{\lambda}$, around  the orthogonal axes. The angles of permissible 
rotations will be determined throughwith a special constraint imposed
upon distortion transformations, namely, a sum of distortions
of corresponding basis vectors $O_{\lambda}$ and 
$\sigma_{\beta}$ has to be zero at given $\lambda$:
\begin{equation}
\label{R52}
<O_{(\lambda\alpha)},O_{\tau}>_{\tau \neq \lambda}+\frac{1}{2}
\varepsilon_{\alpha\beta\gamma}\frac{<\sigma_{(\lambda\beta)},\sigma_{\gamma}>}
{<\sigma_{(\lambda\beta)},\sigma_{\beta}>}=0,
\end{equation}
where $\varepsilon_{\alpha\beta\gamma}$ is an antisymmetric unit tensor.
There up on $\tan\theta_{(\lambda\alpha)}=-\kappa a_{(\lambda\alpha)}$,
where $\theta_{(\lambda\alpha)}$ is the particular rotation around the axis
$\sigma_{\alpha}$ of $R^{3}_{\lambda}$. Inasmuch as the $R$
should be independent of sequence of rotation axes, then  it implies
the mean value $R=\displaystyle \frac{1}{6} \sum_{i \neq j \neq k}
R^{(ijk)}$, where $R^{(ijk)}$ the matrix of rotations carried out 
in sequence $(ijk)$ $(i,j,k=1,2,3)$. As it was seen at the outset the
field $a_{(\lambda\alpha)}$ was generated by the distortion of basis
pseudo-vector $O_{\lambda}$, when the distortion of $\sigma_{\alpha}$
has followed from eq(5.2).\\
Certainly, the whole theory outlined in sections (1-4) will then hold
provided we simply replace each single index $\mu$ of variables by the pair
$(\lambda\alpha)$ and so on. Following to standard rules, next we construct
the diffeomorphism $\eta^{(\lambda\alpha)}(\eta^{(\tau\beta)}_{f}):
G(2.3) \rightarrow G(23)$ and introduce the action eq.(4.1) for the fields.
In the sequel, a transition from six-dimensional curved manifold $G(23)$
to four-dimensional Riemannian geometry $R^{4}$ is straightforward by
making use of reduction of three time-components $e_{0\alpha}=
\displaystyle \frac{1}{\sqrt{2}}(e_{(+\alpha)}+e_{(-\alpha)})$ of basis 
six-vectors $e_{(\lambda \alpha)}$ to single $e_{0}$ in fixed
universal direction. Actually, since Lagrangian of fields on $R^{4}$
is a function of scalars, namely, $A_{(\lambda \alpha)}B^{(\lambda \alpha)}=
A_{0 \alpha}B^{0 \alpha}+A_{\alpha}B^{\alpha}$, so taking into account that
$A_{0 \alpha}B^{0 \alpha}=A_{0 \alpha}<e^{0\alpha},e^{0\beta}>B_{0 \beta}
=A_{0}<e^{0},e^{0}>B_{0}=A_{0}B^{0}$, one readily may perform a
required transition.
The gravitation field equation is written
\begin{equation}
\label{R53}
\partial_{f}^{(\tau \beta)}\partial^{f}_{(\tau \beta)}
a^{(\lambda\alpha)} -(1-\zeta^{-1}_{0})
\partial_{f}^{(\lambda \alpha)}\partial^{f}_{(\tau \beta)}a^{(\tau \beta)}
=-\frac{1}{2}\sqrt{-g}\frac{\partial 
g^{(\tau \beta)(\mu \gamma)}}{\partial a_{(\lambda\alpha)}}
T_{(\tau \beta)(\mu \gamma)}.
\end{equation}
To render our discussion more transparent, below we consider in detail 
a solution of spherical-symmetric static gravitational field
$a_{(+\alpha)}=a_{(-\alpha)}=\displaystyle \frac{1}{\sqrt{2}}a_{0\alpha}
(r_{f})$. So, $\theta_{(+\alpha)}=\theta_{(-\alpha)} =-\arctan 
(\displaystyle \frac{\kappa}{\sqrt{2}}a_{0\alpha})$ and
$\sigma_{(+\alpha)}=\sigma_{(-\alpha)}$. It is convenient to make use of
spherical coordinates $\sigma_{(+1)}=\sigma_{r}$, $\sigma_{(+2)}=
\sigma_{\theta}$, $\sigma_{(+3)}=\sigma_{\varphi}$. The transition
$G(2.3)\rightarrow M^{4}$ is performed by choosing in $T_{f}^{3}$ the
universal direction along radius-vector: $x_{f}^{0r}=t_{f}$, 
$x_{f}^{0\theta}=x_{f}^{0\varphi}=0$. Then, from eq.(5.1) one gets
\begin{equation}
\label{R54}
e_{0}=D^{0}_{0}e^{0}_{0},\quad
e_{r}=D^{r}_{r}e^{0}_{r},\quad
e_{\theta}=e^{0}_{\theta},\quad
e_{\varphi}=e^{0}_{\varphi},
\end{equation}
provided $D=C\otimes I$ with components 
$D^{0}_{0}=1+\displaystyle \frac{\kappa}{\sqrt{2}}a_{0}$,
$D^{r}_{r}=1-\displaystyle \frac{\kappa}{\sqrt{2}}a_{0}$,
$a_{0}\equiv a_{0 r}$, where
$e_{0}^{0}=\xi_{0}\otimes \sigma_{r},\quad
e^{0}_{r}=\xi\otimes \sigma_{r},\quad
e^{0}_{\theta}=\xi\otimes \sigma_{\theta},\quad
e^{0}_{\varphi}=\xi\otimes \sigma_{\varphi}$, 
$\xi_{0}=\displaystyle \frac{1}{\sqrt{2}}(O_{+}+O_{-})$
and $\xi=\displaystyle \frac{1}{\sqrt{2}}(O_{+}-O_{-})$.
The coordinates $x^{\mu}(t,r,\theta,\varphi)$ implying 
$x^{\mu}(x^{l}_{f}):M^{4}\rightarrow R^{4}$ exist in the whole region 
$p^{-1}(U)\in R^{4}$
\begin{equation}
\label{R55}
\frac{\partial x^{\mu}}{\partial x^{l}_{f}}=\psi^{\mu}_{l}=\frac{1}{2}
(D^{\mu}_{l}+\omega^{m}_{l}D^{\mu}_{m}),
\end{equation}
where according to eq.(5.4) one has $x^{0r}=t$, $x^{0\theta}=x^{0\varphi}
=0$. A straightforward calculation gives non-vanishing components
\begin{equation}
\label{R56}
\begin{array}{l}
\psi^{0}_{0}=\displaystyle \frac{1}{2}D^{0}_{0},\quad
\psi^{0}_{1}=\displaystyle \frac{1}{2}t_{f}\partial^{f}_{r}D^{0}_{0},\quad
\psi^{1}_{1}=D^{r}_{r},\quad
\psi^{2}_{2}=\psi^{3}_{3}=0,\\
\omega^{1}_{1}=\omega^{2}_{2}=\omega^{3}_{3}=1, \quad
\omega^{0}_{1}=t_{f}\partial^{f}_{r}D^{0}_{0}.
\end{array}
\end{equation}
Although $\partial_{r}\psi^{0}_{0}=\Gamma^{0}_{01}\psi^{0}_{0}$ and
$\partial_{r}\psi^{1}_{1}=\Gamma^{1}_{11}\psi^{1}_{1}$, but
\begin{equation}
\label{R57}
\partial_{t}\psi^{0}_{1}=\psi^{0}_{0}\partial_{r}^{f}D^{0}_{0}\neq
\Gamma^{0}_{10}\psi^{0}_{0}=
\psi^{0}_{0}\partial_{r}\ln D^{0}_{0},
\end{equation}
where according to eq.(5.4), the non-vanishing components of Christoffel 
symbol are written $\Gamma^{0}_{01}=\displaystyle \frac{1}{2}
g^{00}\partial_{r} g_{00}$,
$\Gamma^{1}_{00}=-\displaystyle \frac{1}{2}
g^{11}\partial_{r} g_{00}$,
$\Gamma^{1}_{11}=\displaystyle \frac{1}{2}
g^{11}\partial_{r} g_{11}$. So, the condition eq.(2.13) holds, namely,
the curvature of $R^{4}$ is not zero. The curved space $R^{4}$ has the 
group of motions $SO(3)$ with two- dimensional space-like orbits $S^{2}$
where the standard coordinates are $\theta$ and $\varphi$. The stationary 
subgroup of $SO(3)$ acts isotropically upon the tangent space at the point of 
sphere $S^{2}$ of radius $r$. So, the bundle $p:R^{4} \rightarrow R^{2}$
has the fiber $S^{2}=p^{-1}(x)$, $x\in R^{4}$ with a trivial connection on 
it, where $R^{2}$ is the factor-space $R^{4}/SO(3)$. In outside of the 
distribution of matter with the total mass $M$, the eq.(5.3) in 
Feynman gauge reduced to $\nabla^{2}_{f}a_{0}=0$, which has the solution
$\displaystyle \frac{\kappa}{\sqrt{2}}a_{0}=-
\displaystyle \frac{GM}{r_{f}}=-
\displaystyle \frac{r_{g}}{2r_{f}}$ (see eq.(4.7)).
So, the line element is written
\begin{equation}
\label{R58}
d\,s^{2}=(1-\frac{r_{g}}{2r_{f}})^{2}dt^{2}-
(1+\frac{r_{g}}{2r_{f}})^{2}dr^{2}-r^{2}(\sin^{2}\theta d\,\varphi^{2}+
d\,\theta^{2}),
\end{equation}
provided by eq.(5.5) and eq.(5.6).Finally, for example, the explicit form of unitary matrix $U_{R}$ of
gravitation group $G_{R}$ in the case of scalar field reads
\begin{equation}
\label{R510}
U_{R}=R'U_{f}R^{+}=e^{\displaystyle {-i[t_{f}\partial^{f}_{r}
\theta_{f}(r_{f})+ \theta_{f}(r_{f})]}},
\end{equation}
where
\begin{equation}
\label{R511}
R=R_{f}=e^{-\displaystyle \frac{igr_{g}}{2r_{f}}t_{f}},\quad
g(a'_{0}-a_{0})=\partial^{f}_{r}\theta_{f}(r_{f}).
\end{equation}
A Lagrangian of charged scalar shadow field eq.(3.4) is in the form
\begin{equation}
\label{R512}
L_{f}(x_{f})=\frac{1}{2}\left[ \left( \frac{7}{8}\right )^{2} 
(D_{l}\varphi_{f})^{*}
D_{l}\varphi_{f} - m^{2}\varphi_{f}^{*}\varphi_{f} \right],
\end{equation}
where, according to eq.(2.11) and eq.(5.6), one has $J_{\Psi}=
\displaystyle \frac{1}{2}(1+\|\omega^{m}_{l}\|)=
\displaystyle \frac{1}{2}$, $S=\displaystyle \frac{7}{8}$.
\section{Distortion of Flat Manifold $G(2.2.3)$ at $G=U^{loc}(1)$}
Following to [4-8], the foregoing theory can be readily generalized for the 
distortion of 12-dimensional flat manifold $G(2.2.3)={}^{*}R^{2}
\otimes {}^{*}R^{2} \otimes R^{3}=\G1_{\eta}(2.3)\oplus
\G1_{u}(2.3)=\displaystyle\sum^{2}_{\lambda,\mu=1} 
\oplus R^{3}_{\lambda \mu}=
{\R_{x}}^{3}\oplus 
{\T_{x}}^{3}\oplus
{\R_{u}}^{3}\oplus 
{\T_{u}}^{3}$ with corresponding 
basis vectors $e_{(\lambda,\mu,\alpha)}=O_{\lambda,\mu}\otimes
\sigma_{\alpha} \subset G(2.2.3)$, 
${\e1_{i}}^{0}_{(\lambda\alpha)}={\O1_{i}}_{\lambda}\otimes 
\sigma_{\alpha}
\subset \G1_{i}(2.3)$, where 
${\O1_{i}}_{+}=
\displaystyle \frac{1}{\sqrt{2}}(O_{1,1} +\varepsilon_{i} O_{2,1})$,
${\O1_{i}}_{-}=
\displaystyle \frac{1}{\sqrt{2}}(O_{1,2} +\varepsilon_{i} O_{2,2})$,
$\varepsilon_{\eta}=1$, $\varepsilon_{u}=-1$ and 
$<O_{\lambda,\mu},O_{\tau,\nu}>={}^{*}\delta_{\lambda \tau}
{}^{*}\delta_{\mu \nu}$ $(\lambda,\mu, \tau, \nu=1,2)$,
$<\sigma_{\alpha},\sigma_{\beta}>=\delta_{\alpha \beta}$. There up on
$<{\O1_{i}}_{\lambda},{\O1_{i}}_{\tau}>=
\varepsilon_{i}\delta_{ij}{}^{*}\delta_{\lambda \tau}$.
The massless gauge field of distortion $a_{(\lambda,\mu,\alpha)}(\zeta_{f})$ 
with the values in Lie algebra of $U^{loc}(1)$ is a local form of
expression of connection in principle bundle $p:E\rightarrow G(2.2.3)$
with a structure group $U^{loc}(1)$. Collection of matter fields
$\Phi_{f}(\zeta_{f})$ are the sections of vector bundles associated with 
$U^{loc}(1)$ by reflection $\Phi_{f}:G(2.2.3)\rightarrow E$ 
that $p \Phi_{f}(\zeta_{f})=\zeta_{f}$, where the coordinates $
\zeta^{(\lambda,\mu,\alpha)}$ exist in the whole region $p^{-1}(U^{(f)})\in
G(2.2.3)$. Outlined theory will then hold, while each pair
of indices $(\lambda \alpha)$ will be replaced by the 
$(\lambda,\mu, \alpha)$. So
the basis $e_{(\tau,\nu,\beta)}$ transformed
\begin{equation}
\label{R61}
e_{(\lambda\mu\alpha)}=D_{(\lambda\mu\alpha)}^{(\tau,\nu,\beta)}
e_{(\tau,\nu,\beta)},
\end{equation}
provided $D=C\otimes R$, where $O_{(\lambda\mu\alpha)}=
C_{(\lambda\mu\alpha)}^{\tau,\nu} O_{\tau,\nu}$,
$\sigma_{(\lambda\mu\alpha)}=
R_{(\lambda\mu\alpha)}^{\beta} \sigma_{\beta}$. The matrices $C$
generate the group of distortion transformations of bi-pseudo-vectors
$O_{\tau,\nu}$: $C_{(\lambda\mu\alpha)}^{\tau,\nu} =
\delta^{\tau}_{\lambda}\delta^{\nu}_{\mu} +\kappa a_{(\lambda,\mu,\alpha)}
{}^{*}\delta^{\tau}_{\lambda}{}^{*}\delta^{\nu}_{\mu}$, but the 
matrices $R$ are the elements of the group $SO(3)_{\lambda \mu}$ of
ordinary rotations of the planes of corresponding basis vectors of
$R^{3}_{\lambda\mu}$. A special constraint eq.(5.2) holds for basis
vectors of each $R^{3}_{\lambda\mu}$. Thus, the gauge field
$a_{(\lambda,\mu,\alpha)}$ was generated by the distortion 
of bi-pseudo-vectors $O_{\tau,\nu}$.  While the rotation transformations 
$R$ follow due to eq.(5.2) and $R$ implies, as usual, the mean value
with respect to sequence of rotation axes. The angles of permissible rotations 
are $\tan \theta_{(\lambda,\mu,\alpha)}=-\kappa a_{(\lambda,\mu,\alpha)}
(\zeta_{f})$. The action eq.(4.1) now reads
\begin{equation}
\label{R62}
S=S_{a_{f}} + S_{\Phi}=\int L_{a_{f}}(\zeta_{f})\,d\zeta^{(1,1,1)}\wedge
\cdots \wedge d\zeta^{(2,2,3)}+
\int \sqrt{g}L_{\Phi}(\zeta)\,d\zeta^{(111)}\wedge
\cdots \wedge d\zeta^{(223)},
\end{equation}
where $g$ is the determinant of metric tensor on $G(223)$ and
$\zeta^{(\lambda\mu\alpha)}(\zeta^{(\tau,\nu,\beta)}):G(2.2.3)
\rightarrow G(223)$ was constructed according to eq.(2.8)-eq.(2.10),
provided 
\begin{equation}
\label{R63}
\frac{\partial \zeta^{(\lambda\mu\alpha)}}{\partial \zeta^{(\tau,\nu,\beta)}}
=\psi^{(\lambda\mu\alpha)}_{(\tau,\nu,\beta)}=\frac{1}{2} \left(
D^{(\lambda\mu\alpha)}_{(\tau,\nu,\beta)}+
\omega^{(\rho,\omega,\gamma)}_{(\tau,\nu,\beta)}
D^{(\lambda\mu\alpha)}_{(\rho,\omega,\gamma)}
\right).
\end{equation}
Right through the variational calculations one infers the field equations
eq.(4.2), while
\begin{equation}
\label{R64}
\begin{array}{l}
\partial^{(\tau,\nu,\beta)}
\partial_{(\tau,\nu,\beta)}
a^{(\lambda,\mu,\alpha)} - (1-\zeta^{-1}_{0})\partial
^{(\lambda,\mu,\alpha)}\partial_{(\tau,\nu,\beta)}a^{(\tau,\nu,\beta)}=
J^{(\lambda,\mu,\alpha)}=\\
=-\displaystyle \frac{1}{2}\sqrt{g}\displaystyle \frac{g
^{(\tau\nu\beta)(\rho\omega\gamma)}}{\partial a_{(\lambda,\mu,\alpha)}}
T_{(\tau\nu\beta)(\rho\omega\gamma)},
\end{array}
\end{equation}
provided
\begin{equation}
\label{R65}
T_{(\tau\nu\beta)(\rho\omega\gamma)}=\frac{2\delta(\sqrt{g}L_{\Phi})}
{\sqrt{g}\delta g^{(\tau\nu\beta)(\rho\omega\gamma)}}.
\end{equation}
The curvature of manifold $\G1_{i}(2.3) \rightarrow 
\G1_{i}(23)$ (sec. 5), which leads to 
four-dimensional Riemannian
geometry $R^{4}$, is a familiar distortion 
\begin{equation}
\label{R66}
a_{(1,1,\alpha)}=a_{(2,1,\alpha)}\equiv \frac{1}{\sqrt{2}}{\a_{\eta}}
_{(+\alpha)},\quad
a_{(1,2,\alpha)}=a_{(2,2,\alpha)}\equiv \frac{1}{\sqrt{2}}{\a_{\eta}}
_{(-\alpha)},
\end{equation}
when $\sigma_{(11\alpha)}=\sigma_{(21\alpha)}\equiv 
\sigma_{(+\alpha)}$, $\sigma_{(12\alpha)}=\sigma_{(22\alpha)}\equiv 
\sigma_{(-\alpha)}$. Hence
${\e1_{i}}
_{(\lambda\alpha)}={\O1_{i}}_{(\lambda\alpha)}\otimes 
\sigma_{(\lambda\alpha)}$, where 
${\O1_{i}}_{(\lambda\alpha)}={\O1_{i}}_{\lambda} +
\displaystyle \frac{\kappa}{\sqrt{2}}\varepsilon_{i}
{\a_{\eta}}_{(\lambda\alpha)}{}^{*}\delta^{\tau}_{\lambda}
{\O1_{i}}_{\tau}$. 
In the aftermath $G(223) = \G1_{\eta}(23)
\oplus \G1_{u}(23)$. 
The other important case of inner-distortion
\begin{equation}
\label{R67}
a_{(1,1,\alpha)}=-a_{(2,1,\alpha)}\equiv \frac{1}{\sqrt{2}}{\a_{u}}
_{(+\alpha)},\quad
a_{(1,2,\alpha)}=-a_{(2,2,\alpha)}\equiv \frac{1}{\sqrt{2}}{\a_{u}}
_{(-\alpha)}
\end{equation}
leads to $\sigma_{(11\alpha)}=-\sigma_{(21\alpha)}\equiv 
\sigma_{(+\alpha)}$, $\sigma_{(12\alpha)}=-\sigma_{(22\alpha)}\equiv 
\sigma_{(-\alpha)}$. Hence,
${\O1_{i}}_{(\lambda\alpha)}=
{\O1_{i}}_{\lambda} +
\displaystyle \frac{\kappa}{\sqrt{2}}\varepsilon_{i}
{\a_{u}}_{(\lambda\alpha)}{}^{*}\delta^{\tau}_{\lambda}
{\O1_{i}}_{\tau}$, 
where the ${\e1_{i}}
_{(\lambda\alpha)}={\O1_{i}}_{(\lambda\alpha)}\otimes 
\sigma_{(\lambda\alpha)}$ is the basis in inner-distorted manifold
$\G1_{i}(23)$.
\section{Discussion and Conclusions}
\label{Dis}
If we collect together the results just established we have finally arrived
at quite promising, if not entirely satisfactory, proposal of generating
the group of gravitation by hidden local internal symmetries. While, under
the reflection of shadow fields from Minkowski flat space to 
Riemannian, one framed the idea of general gauge principle of local internal
symmetries G into requirement of invariance of physical system of reflected
fields on $R^{4}$ under the Lie group of gravitation $G_{R}$ of local
gauge transformations generated by G. This yields the invariance under
wider group of arbitrary curvilinear coordinate transformations in $R^{4}$.
While the energy~-momentum conservation laws are well~-defined.
The fascinating prospect emerged for resolving or mitigating a shortage of 
controversial problems of gravitation including its quantization by 
exploiting whole advantage of field theory in terms of flat space.
In the aftermath, one may carry out an inverse reflection into $R^{4}$
whenever it will be needed.
It was proved that only gravitational attraction exists. One may easily
infer Einstein's equation of gravitation, but with strong difference 
at the vital point of well~-defined energy~-momentum tensor of gravitational
field and conservation laws. Nevertheless, for our part we prefer 
standard gauge invariant Lagrangian eq.(4.11), while the functions
eq.(2.3) ought to be defined.
We considered the gravitational interaction with
hidden Abelian group $G=U^{loc}(1)$ with the base $G(2.3)$
of principle bundle as well as
the distortion of manifold $G(2.2.3)$, which yields both the 
curvature and inner-distortion of space and time.
However, we believe we have made good headway by presenting a reasonable 
framework whereby one will be able to verify the basic ideas and illustrate
main features of drastic generalization of standard gauge principle
of local internal symmetries in order to gain new insight into physical 
nature of gravity.
\vskip0.5truecm
\centerline {\large \bf Acknowledgements}
\vskip 1
\baselineskip
\noindent
It is pleasure to express my gratitude to K.Yerknapetian and A.Vardanian for 
support.

\end {document}